\documentclass[aps,prc,letterpaper,11pt,twoside,tightenlines,nofootinbib,showpacs,preprintnumbers]{revtex4-1}
\usepackage{graphicx}
\usepackage[sort&compress]{natbib}
\usepackage{subfigure}
\usepackage{amsmath}
\usepackage{amsfonts}
\usepackage{cancel}
\usepackage{array}

\begin{document}

\arraycolsep1.5pt

\newcommand{\Ima}{\textrm{Im}}
\newcommand{\Rea}{\textrm{Re}}
\newcommand{\mev}{\textrm{ MeV}}
\newcommand{\be}{\begin{equation}}
\newcommand{\ee}{\end{equation}}
\newcommand{\ba}{\begin{eqnarray}}
\newcommand{\ea}{\end{eqnarray}}
\newcommand{\gev}{\textrm{ GeV}}
\newcommand{\nn}{{\nonumber}}
\newcommand{\dtres}{d^{\hspace{0.1mm} 3}\hspace{-0.5mm}}
\newcolumntype{V}{>{\centering\arraybackslash} m{.5\linewidth} }

\title{Limits to the Fixed Center Approximation to Faddeev equations: the case of the $\phi(2170)$.}

\author{A. Mart\'inez Torres}
\affiliation{Yukawa Institute for Theoretical Physics, Kyoto University, Kyoto 606-8502, Japan}

\author{E. J. Garz\'on, E. Oset}
\affiliation{
Departamento de F\'{\i}sica Te\'orica and IFIC, Centro Mixto Universidad de Valencia-CSIC,
Institutos de Investigaci\'on de Paterna, Aptdo. 22085, 46071 Valencia,
Spain}

\author{L. R. Dai}
\affiliation{Department of Physics, Liaoning Normal University, Dalian, 116029, China}

\preprint{YITP-10-95}

\date{\today}

\begin{abstract}
The Fixed Center Approximation to the Faddeev equations (FCA) has been used lately with success in the study of bound systems of three hadrons. It is also important to set the limits of the approach in those problems to prevent proliferation of inaccurate predictions. In this paper we study the case of the $\phi(2170)$, which has been described by means of Faddeev equations as a resonant state of $\phi$ and $K \bar{K}$, and show the problems derived from the use of the FCA in its study. At the same time we also expose the limitations of an alternative approach recently proposed.
\end{abstract}

\maketitle

\section{Introduction}
\label{Intro}
The Faddeev equations to address the interaction of three body systems \cite{Faddeev:1960su} are very simple conceptually and formally, yet very difficult to solve exactly. Most of the work done with Faddeev equations involve approximations, like the use of separable potentials and energy independent kernels. The Alt-Grassberger-Sandras (AGS) approach \cite{Alt:1967fx} follows this line and is widely used. Recently, studies of three hadron systems, two mesons and one baryon \cite{MartinezTorres:2007sr,Khemchandani:2008rk}, or three mesons \cite{torres} have been done by means of a different approach to the Faddeev equations which relies upon the on shell two body scattering matrices. The method involves approximations of a different kind, derived from the observation that the ratio of the diagrams involving four and three interactions are similar to the ratios of diagrams with three and two interactions. It was shown in Refs.~\cite{MartinezTorres:2007sr,Khemchandani:2008rk} that the low lying $1/2^+$ excited baryons, except the Roper $N^*(1440)$ which is certainly a very complex object, stemmed from the interaction of two pseudoscalar mesons and one baryon. Similarly, the study of the $\phi K \bar{K}$ system in Ref.~\cite{torres} showed that the resonance $X(2175)$ was naturally described in terms of those components with the $K \bar{K}$ pair forming mostly an $f_0(980)$ state.

 The discovery of the $X(2175)$ at BABAR \cite{babar1,babar2} with mass $M_X=2175 \pm 10 \pm 15$~MeV and width $\Gamma_X=58 \pm 16 \pm 20$~MeV~\cite{babar1} in the  $e^+e^-\to\phi(1020)\,f_0(980)$ reaction, was followed by its observation at  BES in $J/\Psi \to \eta\,\phi(1020)\,f_0(980)$ decay with $M_X=2186\pm 10 \pm 6$~MeV and $\Gamma_X=65\pm 23\pm 17$~MeV~\cite{bes08}. The Belle Collaboration  has performed the most precise measurements up to now of the reactions $e^+e^-\to \phi(1020) \pi^+\pi^-$ and $e^+e^-\to\phi(1020)f_0(980)$ finding  $M_X=2079\pm 13^{+79}_{-28}$~MeV and $\Gamma_X=192\pm 23^{+25}_{-61}$~MeV~\cite{belle}. The width obtained is larger than in previous measurements but the errors are larger. A combined fit to both BABAR and Belle data on $e^+e^-\to \phi(1020) \pi^+\pi^-$ and $e^+e^-\to\phi(1020)f_0(980)$  has been done in  Ref.~\cite{Shen:2009mr}, with the results of $M_X=2117^{+0.59}_{-0.49}$~MeV and $\Gamma_X=164^{+69}_{-80}$~MeV. The $X(2175)$ has been renamed in the Particle Data Book (PDG) \cite{pdg} as the $\phi(2170)$ and we shall use this nomenclature from here on.

The $\phi(2170)$ is one of the states recently found which does not stand a clean comparison with predictions of conventional quark model states \cite{Zhu:2007wz}. Much theoretical activity has been developed around the $\phi(2170)$ resonance, suggesting it to be a tetraquark~\cite{wang,hosaka,polosa}, or the lightest hybrid $s \bar{s}g$ state~\cite{dingyan1}. Other works point out at the difficulties encountered trying to interpret the state in terms of already known structures \cite{dingyan2,blacky,Coito:2009na}.
  
  One appealing idea to interpret this resonance was given in Ref.~\cite{torres}. The fact that the resonance is seen in its decay into $\phi$ and $f_0(980)$ suggest that the state could be a strongly bound system of $\phi K \bar{K}$, since in chiral unitary theories the  $f_0(980)$ appears as a resonance of the $\pi \pi$ and $K \bar{K}$ channels, mostly the $K \bar{K}$ one \cite{npa,Kaiser:1998fi,ramonet,Markushin:2000fa}. In Ref.~\cite{torres} the Faddeev equations for the $\phi K \bar{K}$, $\phi \pi \pi$ system were used and the resulting structure was a resonant state of that system with energy and widths inside the range given by the experimental ones and where the $K \bar{K}$ pair was strongly correlated around the $f_0(980)$. The Faddeev equations used for this problem relied upon the chiral unitary two body amplitudes evaluated on shell, once it was proved that the unphysical off shell part of the amplitudes cancel exactly with the explicit three body terms provided by the same chiral Lagrangians 
\cite{MartinezTorres:2007sr,Khemchandani:2008rk}\footnote{This holds in the SU(3) limit. In Ref.~\cite{MartinezTorres:2007sr} it was mentioned that it holds for low momentum transfers, but in Ref.~\cite{Khemchandani:2008rk} it was shown that actually this condition is unnecessary.}.

In view of the technical difficulties to solve the full Faddeev equations one might resort to use a different approximation to these equations, and one of them which is technically very easy is the Fixed Center Approximation (FCA). The basic idea is that one has collisions of one particle against a bound cluster of two other particles, which is not much altered by the interaction with the third particle. The FCA has been used in many problems \cite{Chand:1962ec,Barrett:1999cw,Deloff:1999gc,Kamalov:2000iy,Gal:2006cw} and is accepted as an accurate tool in the study of bound systems, when the particle interacting with the cluster is lighter than the others and the cluster is relatively strongly bound. Yet, in some cases where the interacting particle is heavier than the constituents of the cluster one can still get a qualitative picture from the FCA. This is indeed the case for the $N K \bar{K}$ bound system, with the $K \bar{K}$ making the cluster, which was studied in Ref.~\cite{Xie:2010ig} and results in qualitative agreement with those of Faddeev equations of Ref.~\cite{MartinezTorres:2008kh} were found. Thus, there is no universal rule and for this purpose it is worth to study other systems and other conditions to better understand the limits of the FCA.

More recently, FCA has been also used to study the interaction of systems with several $\rho$ mesons \cite{Roca:2010tf} or of one $K^*$ and several $\rho$ mesons \cite{YamagataSekihara:2010qk}, and in the study of the $K \bar{K} N$ \cite{Xie:2010ig}, the $\bar{K} N N$ \cite{Bayar:2011qj} and the $\pi \rho \Delta$ systems \cite{Xie:2011uw}, where the last two particles make the cluster. In the case of the $K \bar{K} N$ state one is fortunate to be able to compare with the full Faddeev results of Ref.~\cite{MartinezTorres:2008kh,MartinezTorres:2010zv}, as well as with the variational calculations of Ref.~\cite{Jido:2008kp}, and the results of the FCA prove to be rather accurate. Similarly the $\bar{K} N N$ system has been studied with the FCA and chiral dynamics and the results are remarkably similar to those obtained in Ref.~\cite{Dote:2008hw} with a variational calculation, or those of Ref.~\cite{Ikeda:2010tk} with Faddeev equations when a kernel incorporating the energy dependence of chiral dynamics is used. In view of this success, it is also important to recall the limits to its application and we do this in the present paper by choosing a particular case, the $\phi K \bar{K}$ system, for which results with the full Faddeev equations are available in Ref.~\cite{torres}.

Recently, a different technical approach has been proposed in Ref.~\cite{AlvarezRuso:2009xn} to study the $\phi K \bar{K}$ system, leading to the $\phi (2170)$, and further work along these lines has been done in Ref.~\cite{AlvarezRuso:2010je}, hinting at a possible resonant state of the $\phi a_0(980)$, for which no trace was found using the Faddeev equations in Ref.~\cite{torres}. We shall discuss this work here and the problems encountered in that approach.

\section{Fixed center approximation formalism to the $\phi f_0(980)$ scattering}
\begin{figure}[h!]
\centering
\includegraphics[width=\textwidth]{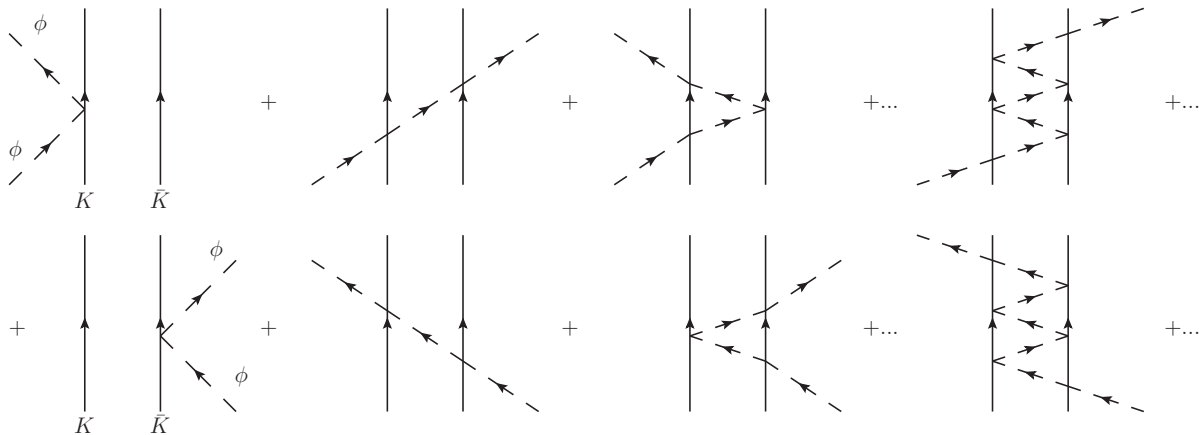}
\caption{Diagrammatic representation of the fixed center approximation to the Faddeev equations.}\label{series}
\end{figure}

By analogy to Refs.~\cite{Chand:1962ec,Barrett:1999cw,Deloff:1999gc,Kamalov:2000iy} we shall study the scattering of a $\phi$ with a molecular state of $K\bar K$. The $\phi$ scatters and rescatters with the $K$, $\bar K$ of the $f_0(980)$ molecule and the $K$, $\bar K$ states are kept unchanged in their wave function of the bound state. Diagrammatically we have the series of terms depicted in Fig.~\ref{series}.
Symbolically these terms can be summed by means of two partition functions $T_1$ (for $\phi K$) and $T_2$ (for $\phi \bar K$), where  $T_1$ sums all the diagrams where the $\phi$ interacts first with the $K$ and $T_2$ those where the $\phi$ interacts first with the $\bar K$. The FCA equations are
\begin{align}
T_1&=t_1+t_1GT_2\nonumber\\
T_2&=t_2+t_2GT_1\label{FCA}\\
T&=T_1+T_2\nonumber
\end{align}
where $T$ is the total $\phi f_0(980)$ $T$-matrix, $G$ is the propagator for the $\phi$ between the $K$, $\bar{K}$ components of the $f_0$ (see Eq.~(\ref{G0}) later on) and $t_1$, $t_2$ stand for the $\phi K$ and $\phi \bar K$ two body $t$-matrices. In order to see the precise meaning and normalizations entering Eq.~(\ref{FCA}) we follow closely the approach of Ref.~\cite{Roca:2010tf}. The formalism is easier here since $\phi K$ and $\phi \bar K$ have the same amplitude and there is only one isospin state $I=1/2$.  We follow Eqs. (19) to (39) of Ref.~\cite{Roca:2010tf} substituting the $f_2(\rho\rho)$ by the $f_0(980) (K\bar K)$ and the external $\rho^+$ by the $\phi$ (see also the approach for the $K^*$ multirho states in Ref.~\cite{YamagataSekihara:2010qk}). The $S$ matrix for single scattering, first diagram of Fig.~\ref{series}, is given by  
 \begin{align}
S^{(1)}&=-it_1 F_{f_0}\Big(\frac{\vec{k}-\vec{k}^\prime}{2}\Big)\frac{1}{{\cal V}^2}
\frac{1}{\sqrt{2\omega_{p_1}}}
\frac{1}{\sqrt{2\omega_{p'_1}}}
\frac{1}{\sqrt{2\omega_k}}
\frac{1}{\sqrt{2\omega_k'}}
(2\pi)^4\,\delta(k+K_{f_0}-k'-K'_{f_0}),
\label{Ssingle}
\end{align}
 where $\cal V$ stands for the volume of a box where we normalize to unity our plane wave states,  $\omega_{p_1}$, $\omega_{p'_1}$, are the energies of the initial and final kaon of the $f_0$ and $\omega_k$, $\omega_{k^\prime}$ the initial and final energy of the $\phi$. The symbols $k$, $k^\prime$, $K_{f_0}$, $K^\prime_{f_0}$ stand for the four momenta of the initial, final $\phi$ and initial, final $f_0$. In Eq.~(\ref{Ssingle}), $F_{f_0}\Big(\frac{\vec{k}-\vec{k}^\prime}{2}\Big)$ is the form factor of the $f_0(980)$ as a bound state of $K\bar K$.
 
The double scattering term of Fig.~\ref{series} gives rise to an $S$ matrix
\begin{align}
S^{(2)}&=-i(2\pi)^4\delta(k+K_{f_0}-k'-K'_{f_0})\frac{1}{{\cal V}^2}
\frac{1}{\sqrt{2\omega_k}} 
\frac{1}{\sqrt{2\omega_k'}}
\frac{1}{\sqrt{2\omega_{p_1}}}
\frac{1}{\sqrt{2\omega_{p'_1}}}
\frac{1}{\sqrt{2\omega_{p_2}}} 
\frac{1}{\sqrt{2\omega_{p'_2}}}\nn\\
&\times \int \frac{d^3q}{(2\pi)^3} 
F_{f_0}\left( \vec{q}-\frac{\vec{k}+\vec{k}^\prime}{2}\right)
\frac{1}{{q^0}^2-\vec{q}\,^2-m_\phi^2+i\epsilon} t_1 t_2.
\label{finalS2}
\end{align}
where $q^0=(s+m_{\phi}^2-M_{f_0}^2)/(2\sqrt s)$ is the $\phi$ energy in the $\phi f_0$ center of mass frame, and $\omega_{p_2}$, $\omega_{p'_2}$, are the energies of the initial and final antikaon.
  
  For scattering at low energies $\vec{k}$, $\vec{k}^\prime$ will be zero and the form factor in the single scattering can be ignored, while the one entering Eq.~(\ref{finalS2}) can be replaced by $F_{f_0} (q)$. The evaluation of this form factor is rendered very easy, and practical for the developments that follow, by using the approach of  Ref.~\cite{Gamermann:2009uq} where a potential in $S$-wave of the type
\begin{equation}
V=v\theta(\Lambda -q)\theta(\Lambda -q')\label{wavep}
\end{equation}
is used in momentum space to obtain a certain bound state (the $f_0$ for instance). In Eq.~(\ref{wavep}) $v$ is a momentum independent function (although it can depend on the energy), and $\Lambda$ a cut off in the modulus of the momenta $q$, $q^\prime$. Following Ref.~\cite{npa} we take 1 GeV for it. As shown in Ref.~\cite{Gamermann:2009uq}, the Quantum Mechanical problem with this potential leads to the same scattering matrix obtained in the chiral unitary approach using the on shell factorization \cite{npa,angels,ollerulf}\footnote{In Ref.~\cite{Gamermann:2009uq} a non relativistic Quantum Mechanical formulation is done but its relativistic extension with the same kernel of Eq.~(\ref{wavep}) to match the field theoretical treatment of the chiral unitary approach is straightforward.}.

By following Ref.~\cite{Gamermann:2009uq} we find the relative wave function of $K\bar K$ in momentum space
\begin{equation}
\langle \vec{p}\,|\psi\rangle 
=v\,\frac{\theta(\Lambda-p)}{E-\omega_K(\vec p_1)-\omega_K(\vec p_2)}\int_{k<\Lambda}
\dtres k \langle \vec{k}|\psi\rangle ,
\end{equation}
which is given in coordinate space by
\begin{equation} 
\langle \vec{x}|\psi\rangle 
= \int \frac{\dtres p}{(2\pi)^{3/2}} e^{i\vec{p}\vec{x}} \langle \vec{p}\,|\psi\rangle . \label{eq26}
\end{equation}
The form factor of the bound $K\bar K$ state is then given by
\begin{equation}
F_{f_0}(q)=\frac{1}{ {\cal N}}
\int_{\substack{p<\Lambda\\|\vec p-\vec q|<\Lambda}}
d^3p\,
\frac{1}{M_{f_0}-2\omega_K(\vec p)}\,
\frac{1}{M_{f_0}-2\omega_K(\vec p-\vec q)},
\label{eq:ff2cutoff}
\end{equation} 
where the normalization factor ${\cal N}$ is
\begin{equation}
{\cal N}=
\int_{p<\Lambda}d^3p
\frac{1}{\left( M_{f_0}-2\omega_K(\vec p) \right)^2}.
\end{equation} 
We will have to face a small technical detail since the field normalization factors in Eqs. (\ref{Ssingle}) and (\ref{finalS2}) are different, and also different than those appearing in the $\phi f_0$ scattering problem in the Mandl and Shaw normalization that we follow \cite{mandl}, which are
\begin{align}
S&=-iT_{\phi f_0}(s)\frac{1}{ {\cal V}^2 }
\frac{1}{\sqrt{2\omega_k}}
\frac{1}{\sqrt{2\omega_{k'}}}
\frac{1}{\sqrt{2\omega_{f_0}}}
\frac{1}{\sqrt{2\omega_{{f_0}'}}}
(2\pi)^4\,\delta(k+K_{f_0}-k'^0-K'_{f_0})
\end{align}
Taking this into account, Eqs.~(\ref{FCA}) can be reformulated in the absence of the form factor in the single scattering as
\begin{align}
T^{(K)}(s)&=\frac{M_{f_0}}{m_K}t_{\phi K}(s^\prime)+\frac{M_{f_0}}{m_K}t_{\phi K}(s^\prime)\tilde{G}_0T^{(\bar K)}\nonumber\\
T^{(\bar K)}(s)&=\frac{M_{f_0}}{m_K}t_{\phi \bar K}(s^\prime)+\frac{M_{f_0}}{m_K}t_{\phi \bar K}(s^\prime)\tilde{G}_0T^{(K)}
\label{FCAeq}\\
T_{\phi f_0}&=T^{(K)}+T^{(\bar K)}, \quad T^{(K)}=T^{(\bar K)}
\end{align}
such that
\begin{equation}
T_{\phi f_0}=2\frac{M_{f_0}}{m_K}t_{\phi K}(s^\prime)\frac{1}{1-\frac{M_{f_0}}{m_K}t_{\phi K}(s^\prime)\tilde{G}_0}\label{TtotnoF}
\end{equation}
 Technically, one should have the ratio of energies instead of that of the masses in Eqs.~(\ref{FCAeq}), but the ratios are nearly the same and we keep this form, as in Ref.~\cite{Roca:2010tf}, for the FCA equations.
 
In the case of the form factor in the single scattering term (the form factor is always present in the rescattering terms in $\tilde{G}_0$) we find
\begin{equation}
T_{\phi f_0}=2\frac{M_{f_0}}{m_K}t_{\phi K}(s^\prime)\left[F_{f_0}\Big(\frac{\vec{k}-\vec{k}^\prime}{2}\Big)-1+\frac{1}{1-\frac{M_{f_0}}{m_K}t_{\phi K}(s^\prime)\tilde{G}_0}\right] \label{TtotF}
\end{equation}
with 
\begin{equation}
\tilde{G}_0\equiv\frac{1}{2M_{f_0}}\int\frac{d^3q}{(2\pi)^3} 
F_{f_0}\Big(\vec{q}-\frac{\vec{k}+\vec{k}^\prime}{2}\Big)
\frac{1}{{q^0}^2-\vec{q}\,^2-m_\phi^2+i\epsilon}.
\label{G0}
\end{equation}
 and
 \begin{align}
 s^\prime&=s^\prime_{\phi K}=(k+p_1)\nonumber\\
 &=(k+p_1+p_2-p_2)^2=s+m^2_K-2\sqrt{s}p^0_2\nonumber\\
 &=m^2_{\phi}+m^2_K+\frac{1}{2}(s-m^2_\phi -m^2_{f_0})
\end{align}
 in the $\phi f_0$ center of mass frame where $s=(k+p_1+p_2)^2$, $p^0_2=E^{CM}_{f_0}/2$.
  
 \section{Field theoretical calculation}
\begin{figure}[h!]
\centering
\includegraphics[width=0.4\textwidth]{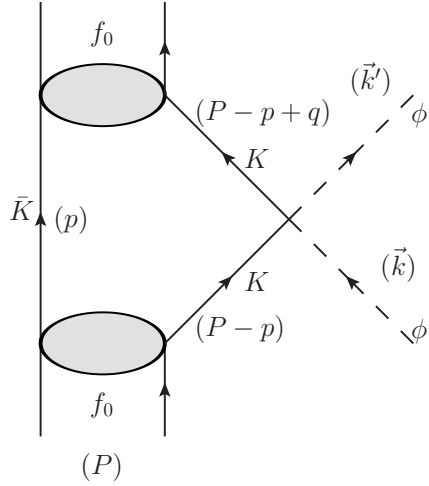}
\caption{Loop diagram considered to evaluate the amplitude in Refs. \cite{AlvarezRuso:2009xn,AlvarezRuso:2010je}. }\label{Luis}
\end{figure}

 In order to make a connection with Refs.~\cite{AlvarezRuso:2009xn,AlvarezRuso:2010je}, let us follow their approach and evaluate the diagram of Fig.~\ref{Luis}. 
  
To facilitate comparison with the results obtained in Ref.~\cite{Gamermann:2009uq}  we use non relativistic propagators for the particles. The extension of both formalism to the relativistic case is straightforward but unnecessary for the result that we shall obtain. The amplitude stemming from the diagram of Fig.~\ref{Luis} is
\begin{align}
-i T^{(FT)}&=(-ig)^2\int \frac{d^4 p}{(2\pi)}(-i \tilde{t}_{\phi K})\frac{i}{p^0-\omega_{\bar K}(\vec{p})+i\epsilon}\frac{i}{P^0-p^0-\omega_K(\vec{p})+i\epsilon}\nonumber\\
&\times \frac{i}{P^0-p^0+q^0-\omega_K(\vec{P}-\vec{p}+\vec{q})+i\epsilon}\label{TFT3}
\end{align}
conveniently regularized, for instance with a cut off, where $q=k-k^\prime$ and $g$ is the coupling of the state $f_0$ to the $K \bar{K}$ channel (residues of the  $K \bar{K} \to K \bar{K}$ amplitude at the pole of the $f_0$ resonance, see Ref.~\cite{Gamermann:2009uq} for details). The amplitude $\tilde{t}_{\phi K}$ in Eq.~(\ref{TFT3}) has a different normalization than the $t_{\phi K}$ used in the rest of the paper to connect with the formalism of Ref.~\cite{Roca:2010tf}. If we take $\vec{P}=0$ (the $f_0$ momentum) and $\vec{q}=0$, the former equation can be written as
\begin{align}
T^{(FT)}&=-ig^2 \tilde{t}_{\phi K}\frac{\partial}{\partial P^0}\int \frac{d^4 p}{(2\pi)}\frac{1}{p^0-\omega_{\bar K}(\vec{p})+i\epsilon}\frac{1}{P^0-p^0-\omega_K(\vec{p})+i\epsilon}\nonumber\\
&=-g^2\tilde{t}_{\phi K}\frac{\partial}{\partial P^0} G(P^0,K\bar K)
\end{align}
where $G$ is the $K\bar K$ propagator or the loop function for $K\bar K$ intermediate states. By using the result of Ref.~\cite{Gamermann:2009uq}, assuming V of Eq.~(\ref{wavep}) independent of energy (see Eq.~(41) of that work, $P^0\equiv E$)
\begin{equation}
g^2=-\Bigg(\frac{d G(E)}{dE}\Bigg)^{-1}_{E=E_{f_0}}\label{g2}
\end{equation}
we obtain the remarkable result for $\vec{q}=0$
\begin{equation}
T^{(FT)}=\tilde{t}_{\phi K}\label{TF}
\end{equation}
which is nothing but the expression of the impulse approximation. The loop of Fig.~\ref{Luis} in field theory implements the impulse approximation used in more conventional approaches.

There is still more to it. Let us perform the $p^0$ integration in the center of mass frame of $\phi f_0$. We have
\begin{equation}
T^{(FT)}=\tilde{t}_{\phi K}g^2\int d^3 p \frac{1}{P^0-\omega_{\bar K}(\vec{p})-\omega_K(\vec{P}-\vec{p})+i\epsilon}\frac{1}{P^0-\omega_{\bar K}(\vec{p})-\omega_K(\vec{P}-\vec{p}+\vec{q})+i\epsilon}\label{TF2}
\end{equation}
where $g^2$, according to Eq.~(\ref {g2}), is the inverse of the $\int d^3 p$ integral of Eq.~(\ref{TF2}) for $\vec{q}=0$. Eq.~(\ref{TF2}) resembles much Eq.~(\ref{eq:ff2cutoff}) that provides the form factor of the $f_0$. One can prove that this corresponds to Eq.~(\ref{eq:ff2cutoff}) using $\vec{q} \rightarrow \vec{q}~^\prime=\frac{\vec{k}-\vec{k}^\prime}{2}$, as obtained in Eq.~(\ref{Ssingle}), by adding to $P^0$ in the second factor of Eq.~(\ref{TF2}) the recoil energy of the final $f_0$, $\vec{q}~^2/2m_{f_0}$. This equivalence is proved in detail in section IV of Ref.~\cite{YamagataSekihara:2010pj}. A different derivation of the same results can be seen in Ref.~\cite{brown} (see also Ref.~\cite{Rogers:2005bt} for a relativistic formulation).

After this discussion about the meaning of the field theoretical approach of Refs.~\cite{AlvarezRuso:2009xn,AlvarezRuso:2010je}, the FCA goes beyond the impulse approximation by taking into account rescattering of the $\phi$ with the $K$ and $\bar K$ of the $f_0$. The rescattering is done zigzagging from one $K$ to the $\bar K$ and viceversa.  Successive scatterings of the $\phi$ on the same particle are forbidden since this leads to diagrams already accounted in the $\phi K$ ($\phi \bar K$) scattering matrix and, thus, one would be double counting. In Refs.~\cite{AlvarezRuso:2009xn,AlvarezRuso:2010je} one is not considering the series implicit in the FCA except for the first term. However, another series of terms is considered as we shall explain in the next section.

\section{Beyond the FCA: excitation of the $f_0$ in intermediate states}
\begin{figure}[h!]
\parbox{0.5\textwidth}{\includegraphics[width=0.5\textwidth]{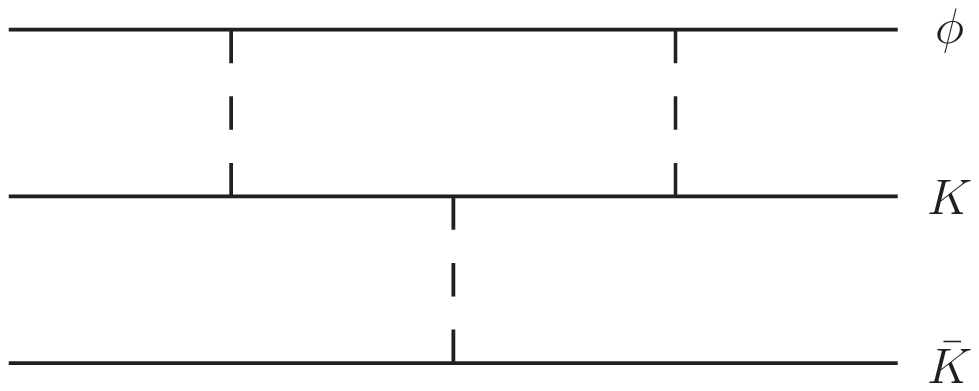}}
\parbox{0.4\textwidth}{\includegraphics[width=0.4\textwidth]{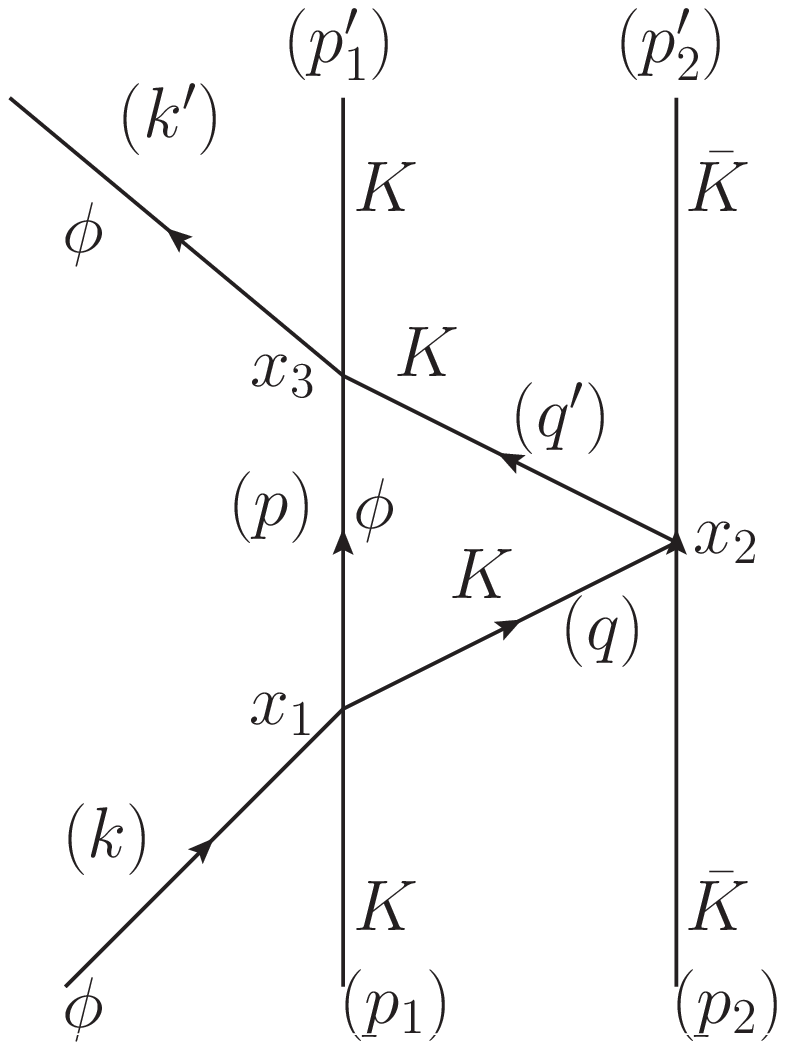}}\\
\parbox{0.5\textwidth}{(a)}
\parbox{0.4\textwidth}{(b)}
\caption{Diagrammatic representation of one of the terms of the Faddeev equations in which the $f_0$ is excited in the intermediate state: a) Typical Faddeev diagram; b) Diagrammatic representation of a) in terms of a field theoretical Feynman diagram.}\label{excit}
\end{figure}

The FCA does not allow for $f_0$ excitation in the intermediate propagation of the $\phi f_0$ states. 
Since for the $\phi(2170)$ one has about $170$ MeV of excitation with respect to the $\phi f_0$ at rest, it looks very unlikely that the $f_0$ is not excited in the intermediate states. An intuitive view can be also obtained by recalling that the $\phi$ has bigger mass than the $K$ and in its collision with the $f_0$ it could easily break this lighter system. To find the explicit answer one resorts to the Faddeev equations and looks for terms where the $f_0$ can be broken in intermediate states. We must have in mind that when one has multiple scattering in the FCA, the $K \bar{K}$ interaction is never used explicitly. Yet, it is implicitly taken into account by using the wave function of the $K \bar{K}$ system. Hence, in the second diagram of Fig.~\ref{series}, the $\phi$ collides with the K, leaving its wave function unchanged, then it propagates and collides with the $\bar{K}$, leaving it also in its original state. This mechanism corresponds to a diagram of the Faddeev expansion with two interactions of the $\phi$ in which the initial and final $K\bar K$ states do not have the possibility of being excited in the intermediate state. We can go one step ahead in the Faddeev series of diagrams and see if the $K \bar{K}$ interaction, in connection with the double scattering of the $\phi$ with the K and the $\bar{K}$, can lead to a break up of the $K \bar{K}$ system once there is sufficient energy for this excitation. This is depicted in the diagram shown in Fig.~\ref{excit}. 
The $S$ matrix for the diagram of Fig.~\ref{excit} is given by
\begin{align}
S^{(3)}&=\frac{1}{\sqrt{2\omega_k}} \frac{1}{\sqrt{2\omega_k'}}\frac{1}{\sqrt{2\omega_{p_1}}}
\frac{1}{\sqrt{2\omega_{p'_1}}}\frac{1}{\sqrt{2\omega_{p_2}}} \frac{1}{\sqrt{2\omega_{p'_2}}}\int d^4 x_1\int d^4 x_2\int d^4 x_3\nonumber\\
&\quad\times\int \frac{d^4 q}{(2\pi)^4}\int \frac{d^4 q^\prime}{(2\pi)^4}\int \frac{d^4 p}{(2\pi)^4} (-i t_{\phi K}(s^\prime))(-i t_{K \bar K}(s^{\prime \prime}))(-i t_{\phi K}(s^\prime))\nonumber\\
&\quad\times\frac{ie^{iq(x_1-x_2)}}{q^2-m^2_{K}+i\epsilon}\,\frac{ie^{iq^{\,\prime}(x_2-x_3)}}{{q^{\,\prime}}^2-m^2_{K}+i\epsilon}\,\frac{ie^{ip(x_1-x_3)}}{p^2-m^2_\phi+i\epsilon}\, e^{i{k^\prime}^0x^0_3}\,e^{i{p^\prime}^0_1x^0_3}\,e^{i{p^\prime}^0_2x^0_2}\,\nonumber\\
&\quad\times e^{-ik^0x^0_1}\,e^{-ip^0_1x^0_1}\,e^{-ip^0_2x^0_2} \frac{1}{\sqrt{\cal V}}e^{-i\vec{k}^\prime\vec{x}_3}\varphi_1(x_3)\varphi_2(x_2)\frac{1}{\sqrt{\cal V}}e^{i\vec{k}\vec{x}_1}\varphi_1(x_1)\varphi_2(x_2)\label{S33}
\end{align}
where 
\begin{align}
s^{\prime\prime}&=(q+p_2)^2=(k+p_1-p+p_2)^2=(k+P-p)^2\nonumber\\
&=s-m^2_\phi-2\sqrt{s}\,\omega_\phi(\vec{p})
\label{ssec}
\end{align}
with $P$ the momentum of the $f_0$.
By performing the $x^0_1$, $x^0_2$, $x^0_3$, $p^0$, ${q^{\,\prime}}^0$, ${p^{\,\prime}}^0$ integrations, making the change of variables
\begin{align}
\vec{x}_1-\vec{x}_2=\vec{r} \quad \quad\quad\vec{x}_3-\vec{x}_2=\vec{r}^{\,\prime}
\end{align}
\begin{align}
\frac{1}{2}(\vec{x}_1+\vec{x}_2)=\vec{R} \quad\quad\quad \frac{1}{2}(\vec{x}_3+\vec{x}_2)=\vec{R}+\frac{\vec{r}^{\,\prime}}{2}-\frac{\vec{r}}{2}
\end{align}
and using that
\begin{align}
\varphi_1(\vec{x}_1)\varphi_2(\vec{x}_2)&=\frac{1}{\sqrt{\cal V}}e^{i\vec{P}(\frac{\vec{x}_1+\vec{x}_2}{2})}\varphi(\vec{r})=\frac{1}{\sqrt{\cal V}}e^{i\vec{P}\vec{R}}\varphi(\vec{r})\\
\varphi^*_1(\vec{x}_3)\varphi^*_2(\vec{x}_2)&=\frac{1}{\sqrt{\cal V}}e^{-i\vec{P}(\frac{\vec{x}_3+\vec{x}_2}{2})}\varphi^*(\vec{r})=\frac{1}{\sqrt{\cal V}}e^{-i\vec{P}^{\,\prime}\vec{R}}
e^{-i\vec{P}^{\,\prime}\frac{\vec{r}^{\,\prime}}{2}}e^{i\vec{P}^{\,\prime}\frac{\vec{r}}{2}}\varphi^*(\vec{r}^{\,\prime})
\label{eqphicomplx}
\end{align}
we can perform the $d^3 x_i$ integrals and the three $\int d^3q$, $\int d^3q^{\,\prime}$, $\int d^3p$ integrals of Eq. (\ref{S33}) and we obtain
\begin{align}
-it^{(3)}&=-i\int\frac{d^3 q^{\,\prime}}{(2\pi)^3}\int\frac{d^3 p}{(2\pi)^3}
t_{\phi K}(s^\prime)t_{K\bar K}(s^{\,\prime \prime})t_{\phi K}(s^\prime)\frac{1}{2\omega_K (\vec{q})}\,
\frac{1}{2\omega_K (\vec{q}^{\,\prime})}\nonumber\\
&\quad\times\frac{1}{2\omega_\phi (\vec{p})}\frac{1}{k^0+p^0_1-\omega_\phi (\vec{p})-\omega_K (\vec{q})+i\epsilon}
\frac{1}{{k^{\,\prime}}^0+{p^{\,\prime}}^0_1-\omega_\phi (\vec{p})-\omega_K (\vec{q}^{\,\prime})+i\epsilon}\nonumber\\
&\quad\times\tilde\varphi\Big(\vec{p}+\vec{q}-\frac{\vec{P}^{\,\prime}}{2}-\frac{\vec{k}+\vec{k}^{\,\prime}}{2}\Big)
\tilde\varphi\Big(\vec{p}+\vec{q}^{\,\prime}-\frac{\vec{P}^{\,\prime}}{2}-\frac{\vec{k}+\vec{k}^{\,\prime}}{2}+\frac{\vec{k}-\vec{k}^{\,\prime}}{2}\Big)\label{S3}\end{align}

As we can see, the amplitude $t_{K\bar K}(s^{\,\prime \prime})$ has appeared with argument $s^{\,\prime  \prime}$, which depends on the integration variables, as shown in Eq.~(\ref{ssec}). In Eq.~(\ref{eqphicomplx}) we have used the complex conjugate of the wave function for formal reasons, although they are real here. For the general case of wave functions and form factors where the system is unbound see section IV of Ref.~\cite{YamagataSekihara:2010pj}.

The amplitude obtained in Eq.~(\ref{S3}) can be compared directly to the double scattering term of Eq.~(\ref{finalS2}), 
\begin{align}
-it^{(2)}&= -i t_{\phi K}(s^\prime)2M_{f_0}\tilde{G}_0t_{\phi K}(s^\prime)
\end{align}
The factor $[t_{\phi K}(s^\prime)]^2$ is the same in both equations and we can remove it for the purpose of comparison of the two amplitudes.

\section{Results}
\begin{figure}[h!]
\centering
\includegraphics[width=0.6\textwidth]{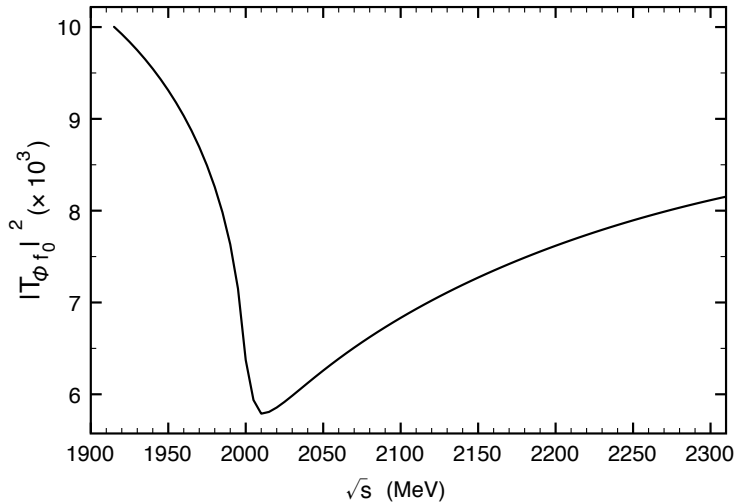}
\caption{Amplitude squared for the $\phi f_0(980)$ interaction without including the form factor $F_{f_0}$. }\label{TtotsqnoF}
\end{figure}

\begin{figure}[h!]
\centering
\includegraphics[width=0.6\textwidth]{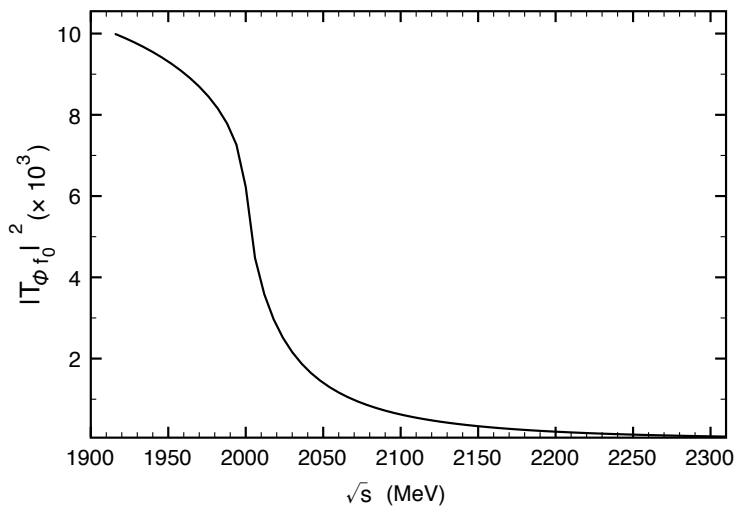}
\caption{Amplitude squared for the $\phi f_0(980)$ interaction including the form factor $F_{f_0}$. }\label{TtotsqF}
\end{figure}

In the calculations we have taken $t_{\phi K}$ from the work of Ref.~\cite{Roca:2005nm}, where the $t$ matrix is obtained in the chiral unitary approach with the $\phi K$ and its coupled channels. We show in Fig.~\ref{TtotsqnoF} the results for the squared $\phi f_0$ amplitude, $|T_{\phi f_0}|^2$, obtained with Eq.~(\ref{TtotnoF}), which omits the form factor in the impulse approximation term, and neglecting the $(\vec{k}+\vec{k}^\prime)/2$ in the argument of the form factor in the $\tilde{G}_0$ function. We can see that the $|T_{\phi f_0}|^2$ goes down when approaching the threshold of $\phi f_0$ and then rises again. We find no sign that there should be a peak around $2170$ MeV. On the other hand, the form factor is here very important since $\vec{k}$ is quite large, of the order of $420$ MeV/c for $\sqrt{s}=2170$ MeV. In the next step we take into account the form factor but project the amplitude over $S$-wave by integrating over $\int d\Omega(\hat {k^\prime})$ the form factor, both in the single scattering term and in $\tilde{G}_0$. The results can be seen in Fig.~\ref{TtotsqF}. We see that the net effect of the form factor has been a drastic reduction of $|T_{\phi f_0}|^2$ beyond the threshold. Below the threshold we have taken $\vec{k}=\vec{k}^\prime=0$ as usually done in these calculations. This is based on the fact that, even if the $\phi K \bar{K}$ system were bound, there is a distribution of real momenta in the wave function, but the momenta are small. One can take a different approach and extrapolate the formula of the form factor below threshold introducing purely imaginary $\phi$ momenta. While one can debate which approach is more physical, it is irrelevant in the present case where we look at the behavior above threshold.

\begin{figure}
\centering
\includegraphics[width=0.6\textwidth]{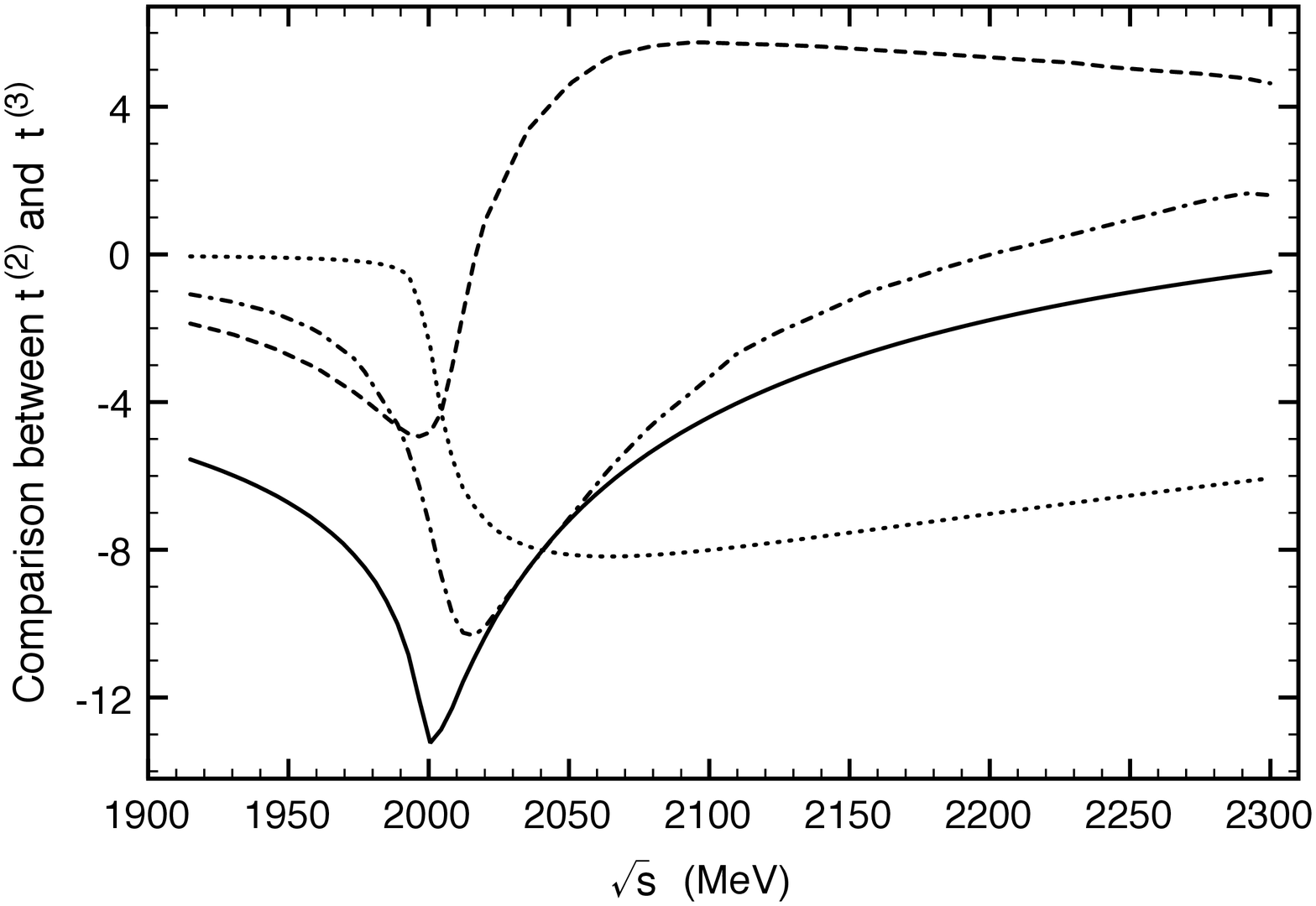}
\includegraphics[width=0.6\textwidth]{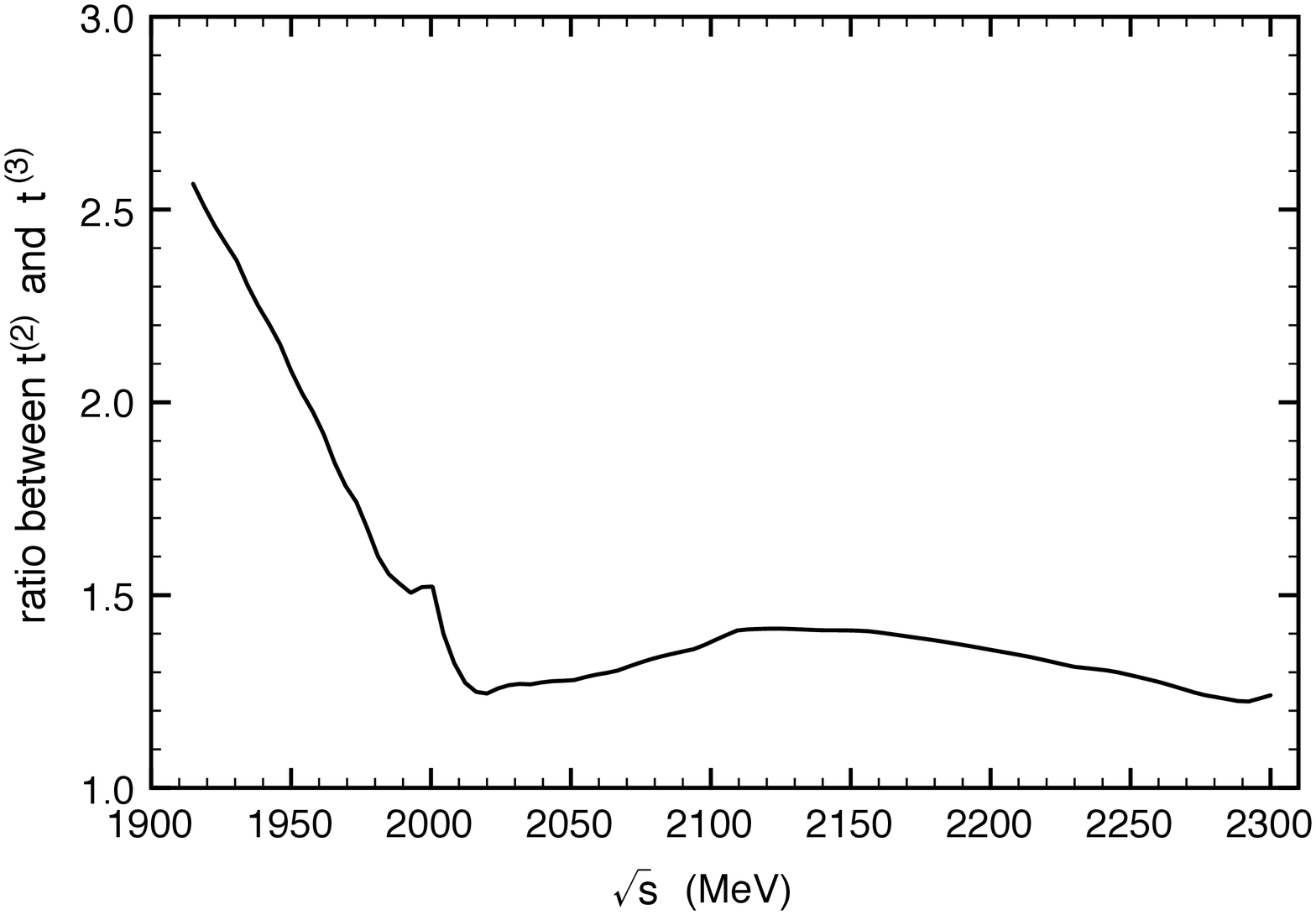}
\caption{Comparison of the amplitudes $t^{(2)}$ and $t^{(3)}$ . (Up) The solid  and dotted line correspond to the real and imaginary part, respectively, of the $t^{(2)}$
amplitude, while the dashed and dash-dotted line are the real and imaginary part, respectively, of the $t^{(3)}$ amplitude, both divided by 
$t_{\phi K}(s^\prime)t_{\phi K}(s^\prime)$, in units of MeV. (Down) Ratio of the modulus of the amplitudes
$t^{(2)}$ and $t^{(3)}$. }\label{compa}
\end{figure}

Once again we can see that there is no trace of a peak around $\sqrt{s}=2170$ MeV when we include the form factor. This is in contrast with the clear peak seen for the $|T_{\phi f_0}|^2$ with the full Faddeev equations, as seen in Fig.1 of Ref.~\cite{torres}.
After this is done, we proceed to evaluate the contribution from the diagram of Fig.~\ref{excit}. In Fig.~\ref{compa} (up) we compare the contribution of this diagram, which now allows for $f_0$ excitation, with the FCA double scattering. In  Fig.~\ref{compa} (down) we show the ratio of the two body term of the FCA to the three body diagram of Fig.~\ref{excit}. We should note that for the case of the two body FCA we have two possibilities, when one starts the $\phi$ interaction from the $K$ or from the $\bar K$, but in the case of Fig.~\ref{excit} we have four possibilities, where the $\phi$ interacts at the beginning and at the end with either the $K$ or the $\bar K$. The relative factor of two in the three body amplitude is incorporated in the figure.

What one can see in Fig.~\ref{compa} is that the three body diagram of Fig.~\ref{excit} with intermediate $f_0$ excitation is of the same size around $\sqrt s =$ 2170 MeV than the double scattering term of the FCA, which has no intermediate $f_0$ excitation. We can see in the figure, comparing real and imaginary parts, that the two mechanisms have amplitudes with opposite signs for the real parts in the region of interest to us. Their simultaneous consideration would change drastically the results obtained from the double scattering contribution of the FCA alone. Yet, when one evaluates the Faddeev integral equations one does not know how much contribution one obtains from such mechanism and this is an information that the present work has provided for the case of the $\phi f_0$ interaction, which should serve as reference for other possible cases where there is also plenty of excitation energy available. Note that the fact that in intermediate states of the multiple scattering the $K \bar{K}$ system is excited does not invalidate that we call this state a $\phi f_0$ system, since asymptotically in the scattering we have indeed $\phi f_0$ in that picture. This is the same as we would have in the $\pi$ elastic scattering with a nucleus in the $\Delta$ region where in intermediate states nucleons can be excited to $\Delta$ states.

At low energies close to threshold, and more clearly below it, the two body FCA amplitude dominates over the excitation term. A more appropriate approach in that region would consist of using a complete set of $K \bar{K}$ states which contained the bound $K \bar{K}$ state and their orthogonal states in the continuum, instead of the basis of plane waves used here. Since the plane waves still have an overlap to the bound wave function, it is logical to think that the use of that alternative basis would give a smaller contribution for the excited states, emphasizing more the role of the ground state of the $f_0$, accounted by the FCA. This numerical finding is in the line with analytical studies of $\pi d$ and $\bar{K} d$ which show that the contribution of the diagram of Fig.~\ref{excit} should vanish at threshold in the limit of $m_K \rightarrow \infty$ \cite{Faldt:1974sm,Baru:2009tx}. At higher energies, where many states can be excited, the use of the set of plane waves for the wave functions becomes progressively more accurate, making our results more realistic in the region of 170 MeV excitation.
  
It is interesting to note that in  Refs.~\cite{AlvarezRuso:2009xn,AlvarezRuso:2010je} the term discussed here of Fig.~\ref{excit} is formally taken into account even if the formalism looks quite different\footnote{We are indebted to J. A. Oller and  L. Alvarez Ruso for clarifying this point to us.}. It is also iterated with multiple steps that have the $f_0~\phi$ in the intermediate states. Yet, the formalism of Refs.~\cite{AlvarezRuso:2009xn,AlvarezRuso:2010je} requires to regularize a loop function for the $f_0~\phi$ propagator, for which one has no input from the derivation of the $K \phi$ and $K \bar{K}$ amplitudes. Thus, one introduces unknown elements at this point, essentially the subtraction constant of a dispersion relation, which is treated as a free parameter in Refs.~\cite{AlvarezRuso:2009xn,AlvarezRuso:2010je}\footnote{One might argue that the Faddeev equations of Refs.~\cite{MartinezTorres:2007sr,Khemchandani:2008rk,torres} also need of an extra regularization factor for the genuine Faddeev loops. In Refs.~\cite{MartinezTorres:2007sr,Khemchandani:2008rk,torres} a cut off of around 1 GeV was used, but only for the purpose of showing the independence of the results with respect to this parameter, since those loops contain three propagators and are convergent without the need of extra cutoffs.}. Furthermore the $K \phi$ interaction, which in the present work is taken from the chiral unitary approach of Ref.~\cite{Roca:2005nm}, is considered in terms of an extra free parameter in Ref.~\cite{AlvarezRuso:2009xn}, which is adjusted to the data of $\phi(2170)$ production. At this point it is important to note that the use of the fitted $\phi K$ in Ref.~\cite{AlvarezRuso:2009xn} results in a $|T_{\phi f_0}|^2$ value at $\sqrt{s} \approx 2170$ MeV about two orders of magnitude bigger than if the $\phi K$ amplitude provided by the chiral unitary approach of Ref.~\cite{Roca:2005nm} was used, which is corroborated by the authors of Ref.~\cite{AlvarezRuso:2009xn} (see Ref.~\cite{jose}). The approach is thus quite different from the Faddeev approach, where the result is fixed before hand from the elementary  $K \phi$ and $K \bar{K}$ scattering amplitudes. Since two free parameters are used in Refs.~\cite{AlvarezRuso:2009xn}, the approach can provide a fit to the data but it is not a predictive scheme for the present problem.

\section{Conclusions}
The fixed center approximation to the Faddeev equations, when two particles are clustered into a bound system, has been often used and proved to be a good approximations in the low energy regime, close to threshold of the constituent particles or below. One particular feature of this approximation is that the cluster of the two particles is not excited in the intermediate states, which can be a good approximation if there is no energy available to excite it. One could hope that it could still be a good approximation if one goes above the threshold, and as usual one does not know the answer until one has checked it. This is what we have done in this paper and we have chosen the case of a problem of current interest, the case of the $\phi(2170)$ particle as a resonant state of the $\phi$ and $f_0(980)$ states. In this case the $\phi ~f_0(980)$ system appears with 170 MeV of excitation, so a priory it looks unlikely that the $f_0$ would not be excited in intermediate states. Yet, it is worth seeing what happens in detail. What we find is that the FCA does not lead to any structure in the region of the $\phi(2170)$, nor close by, unlike by solving the Faddeev equations. On the other hand we determined the contribution of Faddeev diagrams where the $f_0$ is allowed to be excited in the intermediate states, with terms that involve explicitly the $K \bar{K}$ interaction between two $\phi K$ or $\phi \bar{K}$ collisions.  What we found is that this new mechanism is of the same order of magnitude as the $\phi K$ double scattering from the FCA series. Furthermore, we also find that the terms have a strong destructive interference. In view of this, any reliable approach to the problem should take this feature into account. One might think of improving the FCA to include these inelastic excitations, but the problem becomes very involved with the higher iterations and ultimately turns out into a problem far more complicated than the use of the Faddeev equations from the beginning.  Hence, from the practical point of view it does not pay to try to simplify the Faddeev equations in favor of a modified FCA once there is a large excitation energy. 

\section*{Acknowledgments}
We would like to express our must sincere gratitude to Jos\'e Antonio Oller and Luis \'Alvarez Ruso for unveiling their approach and helping us comparing with the results of the present work. This work is partly supported by DGICYT contracts  FIS2006-03438, FPA2007-62777, the Generalitat Valenciana in the program Prometeo and the EU Integrated Infrastructure Initiative Hadron Physics Project  under Grant Agreement n.227431. The work of A. M. T. is supported by the Grant-in-Aid for the Global COE Program ``The Next Generation of Physics, Spun from Universality and Emergence" from the Ministry of Education, Culture, Sports, Science and Technology (MEXT) of Japan. This work is partly supported by ``the National Natural Science Foundation of China (No.10975068) and the Scientific Research Foundation of Liaoning Education Department (No.2009T055)".

\end{document}